\documentclass[aps,prd,preprintnumbers,nofootinbib]{revtex4}
\usepackage[dvips]{graphicx}
\usepackage{bm,latexsym,amsmath,amssymb,amsfonts}

\begin{document}

\title{A no-go on strictly stationary spacetimes in four/higher dimensions}

\author{Tetsuya Shiromizu, Seiju Ohashi and Ryotaku Suzuki}
\affiliation{Department of Physics, Kyoto University, Kyoto 606-8502, Japan}
\begin{abstract}
We show that  strictly stationary spacetimes cannot have non-trivial configurations of 
form fields/complex scalar fields and then the spacetime should be exactly Minkowski 
or anti-deSitter spacetimes depending on the presence of negative cosmological 
constant. That is, self-gravitating complex scalar fields and form fields cannot exist. 
\end{abstract}
\maketitle

\section{Introduction}

Whether the self-gravitating and stationary/static configuration exists or not is fundamental 
issue in general relativity. The famous and elegant Lichnerowicz theorem tells us that the 
vacuum and 
strictly stationary spacetimes should be static \cite{Lich1955}. The phrase ``strict stationarity" means that 
the existence of the timelike Killing vector field in the whole region of the spacetime is assumed 
(no black holes!). Since the total mass of the spacetime is zero, the positive mass theorem shows us that the 
spacetime should be the Minkowski spacetime  \cite{Schoen81, Witten81}.  The similar discussion has 
been extended into asymptotically anti-deSitter spacetimes \cite{BGH}. But, we know that there 
is the static non-trivial solution for the Einstein-Yang-Mills systems \cite{Bartnik88} (See Ref. 
\cite{Winstanley2008} and references therein). Holographic arguments of condensed matters is deeply 
related to the existence of non-trivial self-gravitating configurations in asymptotically anti-deSitter 
spacetimes (See Ref. \cite{Hartnoll:2011fn} for a review). 

Recently, there are interesting new issues on self-gravitating objects and final fate of 
gravitational collapse in asymptotically anti-deSitter spacetimes \cite{Dias:2011at, Bizon:2011gg, 
Dias:2011ss}(See also Ref. \cite{Stotyn2012, Gentle2012}). 
Therein non-stationary numerical solutions with $1$-Killing vector field was found in the 
Einstein-complex scalar system \cite{Dias:2011at}(This is a kind of boson stars. For 
boson stars, see Ref. \cite{Liebling2012} and reference therein). A kind of 
no-go theorem or Lichnerowicz-type theorem is also important. This is because they provide us 
some definite informations about the above issue in an implicit way. 

In this paper we present a no-go theorem for non-trivial self-gravitating configurations 
composed of $p$-form fields/complex scalar fields in strictly stationary spacetimes. We will 
consider asymptotically flat or anti-deSitter spacetimes. This no-go does not contradict with the results of 
recent works \cite{Dias:2011at, Bizon:2011gg, Dias:2011ss} because the spacetimes there are not 
strictly stationary or there are some coupling between gauge fields and scalar fields and so on. 
See Ref. \cite{Gibbons:1990um} for related issues about static configurations. 

The organisation of this paper is as follows. In Sec. II, we discuss the strictly stationary 
spacetimes in four dimensions, and show that the Maxwell field and complex scalar field cannot 
have non-trivial configuration. In Sec. III, we generalised this into higher dimensions with 
$p$-form fields and complex scalar fields. In Appendix A, we present a technical detail. 
In Appendix B, we discuss an alternative argument for asymptotically anti-deSitter spacetimes.

\section{four dimensions}

Bearing the recent work \cite{Dias:2011at} in mind, we consider the following system 
%
\begin{equation}
L=R-\frac{1}{2}F^2-2|\partial \pi|^2-2\Lambda,
\end{equation} 
%
$F$ and $\pi$ are the field strength of the Maxwell field and a complex scalar field, respectively. 
There is no source term for the Maxwell field and no potential of $\pi$. 
The Einstein equations are  
%
\begin{equation}
R_{ab}=F_a^{~c}F_{bc}-\frac{1}{4}g_{ab}F^2+\partial_a \pi \partial_b \pi^*+\partial_a \pi^* \partial_b \pi  + \Lambda g_{ab}. 
\end{equation} 
%
Let us focus on the strictly stationary spacetimes, that is, we assume that there is a timelike 
Killing vector field $k^a$ everywhere. In addition, we assume that the Maxwell field and complex scalar 
field are also stationary, $\mbox \pounds_k F=0,  k^a \partial_a \pi =0$. Then we see 
$\mbox \pounds_k T_{ab}=0$ which is consistent with the spacetime stationarity. 

The twist vector $\omega^a$ is defined by 
%
\begin{equation}
\omega_a=\frac{1}{2}\epsilon_a^{~bcd}k_b \nabla_c k_d. 
\end{equation} 
%
Then, from the definition of $\omega_a$, one can show 
%
\begin{equation}
\nabla_a (\omega^a V^{-4})=0, \label{divtwist}
\end{equation} 
%
where $V^2=-k_ak^a$. 
We introduce the electric and magnetic components of the Maxwell field as 
%
\begin{equation}
E_a = k^b F_{ba}
\end{equation}
and
\begin{equation}
B_a = - \frac{1}{2} \epsilon_{abcd}F^{bc}k^d,
\end{equation}
%
respectively. 
Using $E^a$ and $B^a$, the field strength of the Maxwell field is written as
%
\begin{equation}
V^2F_{ab}=-2k_{[a}E_{b]}+\epsilon_{abcd}k^cB^d.
\end{equation} 
%
The source-free Maxwell equation becomes
%
\begin{equation}
\nabla_{[a}E_{b]}=0,
\end{equation} 
%
%
\begin{equation}
\nabla_{[a}B_{b]}=0,
\end{equation} 
%
%
\begin{equation}
\nabla_a (E^aV^{-2})-2 \omega_a B^a V^{-4}=0
\end{equation} 
%
and
%
\begin{equation}
\nabla_a (B^aV^{-2})+2 \omega_a E^a V^{-4}=0.
\end{equation} 
%
From the first two equations, we see that $E^a$ and $B^a$ have the potentials as 
%
\begin{equation}
E_a=\nabla_a \Phi,~~B_a=\nabla_a \Psi.
\end{equation} 
%
Using the Einstein equations, we can show that 
%
\begin{equation}
\nabla_{[a}\omega_{b]}=B_{[a}E_{b]}
\end{equation} 
%
holds. The right-hand side is the Poynting flux. To show the above, we used the assumption of 
the stationarity of the complex scalar field. Using the fact that the electric and magnetic 
fields are written in terms of the potential for each, we can rewrite the above in several 
different ways, for examples, we have the following typical two equations 
%
\begin{equation}
\nabla_{[a}(\omega_{b]}-\Psi E_{b]})=0 
\end{equation} 
%
and
%
\begin{equation}
\nabla_{[a}(\omega_{b]}+\Phi B_{b]})=0.
\end{equation} 
%
Therefore, the existence of scalar functions are guaranteed as 
%
\begin{equation}
\omega_a-\Psi E_a=\nabla_a U_E \label{Ue}
\end{equation} 
%
and
%
\begin{equation}
\omega_a+\Phi B_a=\nabla_a U_B. \label{Ub}
\end{equation} 
%
Then, using the Maxwell equation, Eqs. (\ref{divtwist}), (\ref{Ue}) and (\ref{Ub}) give us 
%
\begin{equation}
\nabla_a \Bigl(U_E \frac{\omega^a}{V^4}-\frac{\Psi}{2V^2}B^a \Bigr)=\frac{\omega_a \omega^a}{V^4}-\frac{B_a B^a}{2V^2}.
\label{UE}
\end{equation} 
%
and
%
\begin{equation}
\nabla_a \Bigl(U_B \frac{\omega^a}{V^4}-\frac{\Phi}{2V^2}E^a \Bigr)=\frac{\omega_a \omega^a}{V^4}-\frac{E_a E^a}{2V^2}.
\label{UB}
\end{equation} 
%
They correspond to Eq. (6.35) in page 156 of Ref. \cite{Carter} which has 
an error of the sign(For example, see Ref. \cite{Heusler1998}). 

On the other hand, the Einstein equations give us  
%
\begin{equation}
\frac{2}{V^2}R_{ab}k^ak^b=
\nabla_a \Bigl(\frac{\nabla^a V^2}{V^2} \Bigr)  + 4 \frac{\omega_a \omega^a}{V^4}
=-2\Lambda + \frac{E_a E^a+B_a B^a}{V^2}. \label{rkk}
\end{equation} 
%
Note that this corresponds to the Raychaudhuri equation for non-geodesics. 

It is easy to see that Eqs. (\ref{UE}), (\ref{UB}) and (\ref{rkk}) implies 
%
\begin{equation}
\nabla_a \Bigl(\frac{\nabla^a V^2}{V^2}+W^a \Bigr)=-2\Lambda,
\end{equation} 
%
where 
%
\begin{equation}
W^a=2(U_E+U_B)\frac{\omega^a}{V^4}-\frac{\Psi B^a+\Phi E^a}{V^2}.
\end{equation} 
%
Let us first consider the cases with $ \Lambda=0$. Then we see 
%
\begin{equation}
\nabla_a \Bigl(\frac{\nabla^a V^2}{V^2}+W^a \Bigr)=0,\label{vw}
\end{equation} 
%
The space volume integral of the above implies the surface integral. Since $W^a$ does not 
contribute to the surface integral, we see that it will be the total mass and 
then 
%
\begin{equation}
M=0
\end{equation} 
%
(See Appendix A). 
Here note that the positive mass theorem holds because the dominant energy condition is satisfied. 
Thus, the corollary of the positive mass theorem tells us that the spacetime should be the Minkowski 
spacetime \cite{Schoen81, Witten81}. This means that the electro-magnetic fields and complex scalar field vanish. 

Next we consider the cases with $\Lambda <0$. Introducing the vector field, $r^a$, satisfying 
%
\begin{equation}
\nabla_a r^a=-2\Lambda, \label{divr}
\end{equation} 
%
we find 
%
\begin{equation}
\nabla_a \Bigl(\frac{\nabla^a V^2}{V^2}-r^a+W^a \Bigr)=0. \label{surface}
\end{equation} 
%
The volume integral shows us 
%
\begin{equation}
M=0
\end{equation} 
%
again. Then the positive mass theorem implies that the spacetime should be the exact anti-deSitter spacetime 
\cite{Gibbons83, Gibbons83-2}. 
In Appendix A, we discuss the existence of $r^a$ satisfying Eq. (\ref{divr}). However, one may not want to 
introduce this $r^a$. This is possible for a restricted case. In Appendix B, we present an alternative proof for 
asymptotically anti-deSitter spacetimes. Price we  have to pay is that we cannot include the Maxwell field 
in the argument. 

One may wonder if one can extend this result to the cases with  positive cosmological constant, $\Lambda>0$. 
Although there are some efforts \cite{dSPET}, we do not have the positive mass theorem which holds for general 
asymptotically deSitter spacetimes. Thus, we cannot have the same statement with our current result. 

Note that almost of all basic equations presented here was derived in Ref. \cite{Heusler1996}. 
However, these equations was applied to the stationary {\it axisymmetric} black holes to derive the Smarr formula 
and so on. On the other hand, we focused on the strictly stationary spacetimes which are not 
ristricted to be axisymmetric in general and do not contain black holes.

\section{Higher dimensions}

Let us examine the same issue in higher dimensions. The Lagrangian we consider is 
%
\begin{equation}
{\cal L}=R-\frac{1}{p!}H_{(p)}^2-2|\partial \pi|^2-2\Lambda,
\end{equation} 
%
where $H_{(p)}$ is the field strength of a $(p-1)$-form field potential and $\pi$ is a 
complex scalar field. We consider the strictly stationary spacetimes, $p$-form fields and complex 
scalar fields, $\mbox \pounds_k H_{(p)}=0, \mbox \pounds_k \pi=0$. 

The Einstein equations become 
%
\begin{eqnarray}
R_{ab} =  \frac{1}{p!}\Bigl(pH_a^{~c_1 \cdots c_{p-1}}H_{bc_1 \cdots c_{p-1}}-\frac{p-1}{n-2}g_{ab}H_{(p)}^2 \Bigr)
+\partial_a \pi \partial_b \pi^*+\partial_a \pi^* \partial_b \pi
+\frac{2}{n-2}\Lambda g_{ab}.  
\end{eqnarray} 
%
The field equations for source free $p$-form field are 
%
\begin{eqnarray}
\nabla_a H^{a_1 \cdots a_{p-1}a}=0
\end{eqnarray} 
%
and the Bianchi identity. 
Let us decompose the $p$-form field strength into the electric~($E_{a_1 \cdots a_{p-1}}$) and 
magnetic parts~($B_{a_1 \cdots a_{n-p-1}}$) as 
%
\begin{eqnarray}
V^2H_{a_1 \cdots a_{p}} =  -p k_{[a_1}E_{a_2 \dots a_{p}]} +\epsilon_{a_1 \cdots a_{p}a_{p+1}a_{p+2}\cdots a_n}
k^{a_{p+1}}B^{a_{p+2}\cdots a_n}.
\end{eqnarray} 
%
where we define the each component by
%
\begin{eqnarray}
E_{a_1 \dots a_{p-1}}=H_{aa_1 \cdots a_{p-1}}k^a
\end{eqnarray} 
%
and
%
\begin{eqnarray}
B_{a_1 \dots a_{n-p-1}}=\frac{1}{p!(n-p-1)!}\epsilon_{b_1 \cdots b_p c a_1 \cdots a_{n-p-1}}k^cH^{b_1 \cdots b_{p1}}.  
\end{eqnarray} 
%

Here we define the twist tensor $\omega_{a_1 \cdots a_{n-3}}$ as 
%
\begin{eqnarray}
\omega_{a_1 \cdots a_{n-3}}=\alpha \epsilon_{a_1 \cdots a_{n-3}bcd}k^b \nabla^c k^d,
\end{eqnarray} 
%
where $\alpha$ is a constant. From the definition of the twist, it is easy to check that 
%
\begin{eqnarray}
\nabla_{a_{n-3}}\Bigl(\frac{\omega^{a_1 \cdots a_{n-3}}}{V^4} \Bigr)=0 \label{divw}
\end{eqnarray} 
%
holds. From the field equations, we have 
%
\begin{eqnarray}
\nabla_{[a_1} E_{a_2 \cdots a_{p}]}=0
\end{eqnarray} 
%
and
%
\begin{eqnarray}
\nabla_{[a_1} B_{a_2 \cdots a_{n-p}]}=0. 
\end{eqnarray} 
%
Then there are the potentials, that is, 
%
\begin{eqnarray}
E_{a_1 \cdots a_{p-1}}=\nabla_{[a_1} \Phi_{a_2 \cdots a_{p-1}]}
\end{eqnarray} 
%
and
%
\begin{eqnarray}
B_{a_1 \cdots a_{p-1}}=\nabla_{[a_1} \Psi_{a_2 \cdots a_{n-p-1}]}. 
\end{eqnarray} 
%
It is seen from the definition of $E_{a_1 \cdots a_{p-1}}$ and $B_{a_1 \cdots a_{p-1}}$ that 
$k^{a_1} \Phi_{a_1\dots a_{p-2}}= 0$ and $k^{a_1} \Psi_{a_1\dots a_{n-p-2}}=0$ hold. 
The other field equations give us 
%
\begin{eqnarray}
\Phi_{a_1 \cdots a_{p-2}}\nabla_{a_{p-1}} \Bigl(\frac{E^{a_1 \dots a_{p-2}a_{p-1}}}{V^2} \Bigr)
=\alpha^{-1}(-1)^n \omega^{a_1 \cdots a_{p-2}b_1 \cdots b_{n-p-1}}\Phi_{a_1 \cdots a_{p-2}}\frac{B_{b_1 \cdots b_{n-p-1}}}{V^4}
\label{twistB}
\end{eqnarray} 
%
and
%
\begin{eqnarray}
\Psi_{a_1 \cdots a_{n-p-2}}\nabla_{a} \Bigl(\frac{B^{a_1 \dots a_{n-p-2}a}}{V^2} \Bigr)
=-\alpha^{-1}\frac{(-1)^p}{(n-p-1)!(p-1)!} \omega^{b_1 \cdots b_{p-1}a_1 \cdots a_{n-p-2}}\Psi_{a_1 \cdots a_{n-p-2}}
\frac{E_{b_1 \cdots b_{p-1}}}{V^4}. 
\end{eqnarray} 
%
Using the Einstein equations, we can show 
%
\begin{eqnarray}
\alpha^{-1}\epsilon^{abcd_1 \cdots d_{n-3}}\nabla_c \omega_{d_1 \cdots d_{n-3}}
& = & 2(n-3)!(-1)^n (k^aR^b_{~c}-k^bR^a_{~c})k^c \nonumber \\
& = & -\frac{2(-1)^{n+p}(n-3)!}{(p-1)!} \epsilon^{abcd_1 \cdots d_{n-3}}E_{cd_1 \cdots d_{p-2}}B_{d_{p-1}\cdots d_{n-3}}.
\end{eqnarray} 
%
Note that we used the fact of the stationarity of the complex scalar field. 
Then we see that there are the $(n-4)$-forms $U^E$ and $U^B$ satisfying 
%
\begin{eqnarray}
\nabla_{[a_1}U^E_{a_2 \cdots a_{n-3}]}
=\omega_{a_1 \cdots a_{n-3}}-\frac{(-1)^n 2\alpha (n-3)!}{(p-1)!} E_{[a_1 \cdots a_{p-1}}\Psi_{a_p \cdots a_{n-3}]}
\end{eqnarray} 
%
and
%
\begin{eqnarray}
\nabla_{[a_1}U^B_{a_2 \cdots a_{n-3}]}
=\omega_{a_1 \cdots a_{n-3}}+\frac{(-1)^{n+p} 2\alpha (n-3)!}{(p-1)!} \Phi_{[a_1 \cdots a_{p-2}}B_{a_{p-1} \cdots a_{n-3}]},
\end{eqnarray} 
%
respectively. Using $U^{E,B}$ and Eq. (\ref{divw}), we can have the following equations
%
\begin{eqnarray}
\nabla_a \Bigl( U^E_{a_1 \cdots a_{n-4}}\frac{\omega^{a_1 \cdots a_{n-4}a}}{V^4}
-(-1)^p \alpha^2 \beta \frac{\Psi_{a_1 \cdots a_{n-p-2}}B^{a_1 \cdots a_{n-p-2}a}}{V^2}
  \Bigr)
=(-1)^n \Bigl(\frac{\omega^2}{V^4}-\alpha^2 \beta \frac{B^2}{V^2}\Bigr) \label{idE}
\end{eqnarray} 
%
and
%
\begin{eqnarray}
\nabla_a \Bigl( U^B_{a_1 \cdots a_{n-4}}\frac{\omega^{a_1 \cdots a_{n-4}a}}{V^4}
-(-1)^{n+p} \alpha^2 \gamma \frac{\Phi_{a_1 \cdots a_{p-2}}E^{a_1 \cdots a_{p-2}a}}{V^2}
  \Bigr)
=(-1)^n \Bigl(\frac{\omega^2}{V^4}-\alpha^2 \gamma \frac{E^2}{V^2}\Bigr),\label{idB}
\end{eqnarray} 
%
where $\beta=2(n-3)!(n-p-1)!$ and $\gamma=2(n-3)!/(p-1)!$. On the other hand, the Einstein equations 
give us 
%
\begin{eqnarray}
\frac{2}{V^2} R_{ab}k^ak^b=\nabla_a \Bigl( \frac{\nabla^a V^2}{V^2} \Bigr)  + \frac{1}{\alpha^2 (n-3)!} \frac{\omega^2}{V^4}
=\frac{2(n-p-1)}{(p-1)!(n-2)}\frac{E^2}{V^2}+\frac{2(p-1)(n-p-1)!}{n-2}\frac{B^2}{V^2}
-\frac{4}{n-2}\Lambda. \label{ein}
\end{eqnarray} 
%
Then, together with Eqs. (\ref{idE}) and (\ref{idB}), this implies 
%
\begin{eqnarray}
\nabla_a \Bigl( \frac{\nabla^a V^2}{V^2}+X^a \Bigr)=-\frac{4}{n-2}\Lambda, \label{surface2}
\end{eqnarray} 
%
where $X^a$ is defined by 
%
\begin{eqnarray}
X^a& = & \frac{(-1)^n}{\alpha^2(n-2)!}\Bigl( (p-1)U^E_{a_1 \cdots a_{n-4}}+(n-p-1)U^B_{a_1 \cdots a_{n-4}} \Bigr) 
\frac{\omega^{a_1 \cdots a_{n-4}a}}{V^4}\nonumber \\
& & -(-1)^{n+p}\frac{2(p-1)(n-p-1)!}{n-2} 
\frac{\Psi_{a_1 \cdots a_{n-p-2}}B^{a_1 \cdots a_{n-p-2}a}}{V^2}
-(-1)^p \frac{2(n-p-1)}{(p-1)!(n-2)}
\frac{\Phi_{a_1 \cdots a_{p-2}}E^{a_1 \cdots a_{p-2}a}}{V^2}.
\end{eqnarray} 
%
In the similar argument with four dimensional case,  
we can show that the mass vanishes. Then, using the positive mass theorem in higher dimensions
\footnote{If one considers spin manifolds, the positive mass theorem is easily proven as in the four 
dimensional Witten's version \cite{Witten81}}, 
this means that the spacetime is exactly Minkowski/anti-deSitter 
spacetimes depending on the presence of the negative cosmological constant. In any cases, 
the $p$-form fields and complex scalar fields vanish. For asymptotically anti-deSitter case, 
we had to introduce the vector field $r^a$ satisfying 
%
\begin{eqnarray}
\nabla_a r^a=-\frac{4}{n-2}\Lambda. \label{defr}
\end{eqnarray} 
%
The existence of this $r^a$ is discussed in Appendix A. As in the four dimensions, 
the argument which avoids to introduce $r^a$ is given in Appendix B.
 
\section{Summary and discussion}

In this paper we showed that strictly stationary spacetimes with 
$p$-form and complex scalar fields should be Minkowski or anti-deSitter spacetime 
depending on the presence of the negative cosmological constant.  

From our result, we have no room where we have the self-gravitating solution composed of complex 
scalar fields in strictly stationary spacetimes. Therefore, if one wishes to explore new solution, 
one has to think of the set-up which breaks some assumptions imposed here. For example, we can 
find a new configuration which is non-stationary, but has non-stationary $1$-Killing vector field 
\cite{Dias:2011at}.

For asymptotically anti-deSitter spacetimes, we had to introduce a vector $r^a$ to show the 
no-go. As shown in Appendix B, there is a way to avoid this additional treatment for the 
Einstein-complex scalar system. However, it it quite hard to extend this into the cases with 
$p$-form fields. This is left for future works.

\begin{acknowledgments}
 TS is partially supported by 
Grant-in-Aid for Scientific Research from Ministry of Education, Culture, Sports, Science and Technology(MEXT) 
of Japan (No.~21244033). SO is supported by JSPS Grant-in-Aid for Scientific Research (No. 23-855). 
SO and RS are supported by the Grant-in-Aid for the Global COE Program "The
Next Generation of Physics, Spun from Universality and Emergence" from MEXT of Japan. 
The authors also thank the Yukawa Institute for Theoretical Physics at Kyoto University, 
where this work was initiated during the YITP-T-11-08 on "Recent advances in numerical and 
analytical methods for black hole dynamics". 
\end{acknowledgments}

\appendix

\section{The evaluation of boundary term}

In this Appendix, we present the detail of the calculation of the integral of 
Eqs. (\ref{surface}) and (\ref{surface2}). Since the argument for asymptotically flat spacetimes is 
included in asymptotically anti-deSitter cases, we will focus on the latter. 

In asymptotically anti-deSitter spacetimes, the leading behaviour of the metric is 
%
\begin{eqnarray}
ds^2=-V^2dt^2+V^{-2}dr^2+r^2d\Omega_{n-2}^2+\cdots,
\end{eqnarray} 
%
where 
%
\begin{eqnarray}
V^2=1-\frac{2M}{r^{n-3}}-\frac{2}{(n-2)(n-1)}\Lambda r^2.
\end{eqnarray} 
%
Let us consider the vector $r^a$ satisfying 
%
\begin{eqnarray}
\nabla_a r^a=-\frac{4}{n-2}\Lambda. 
\end{eqnarray} 
%
Near the infinity, $r^a$ will be given by 
%
\begin{eqnarray}
r^a \simeq -\frac{4}{(n-2)(n-1)}\Lambda r (\partial_r)^a +\cdots.
\end{eqnarray} 
%
The global existence of $r^a$ is guaranteed as follows. Without loss of generality, 
one can suppose the form of $r^a=\nabla^a \varphi$. Then the above equation becomes 
$\nabla^2 \varphi=-2\Lambda$. We can redefine $\varphi$ so that $-2\Lambda$ is subtracted and then 
$\nabla^2 \tilde \varphi =S$, where $S$ is a non-singular source term. 
The existence of the solution to this is well-known fact in regular Riemannian manifolds. 
Therefore, we can always introduce that $r^a$ in general. 

For the vectors $V^a$ satisfying $V^ak_a=0$, the space-time divergence is written as 
%
\begin{eqnarray}
\nabla_a V^a=\frac{1}{{\sqrt {-g}}} \partial_\mu ({\sqrt {-g}} V^\mu)=\frac{1}{V{\sqrt {q}}} \partial_i (V{\sqrt {q}} V^i)
\end{eqnarray} 
%
where $q$ is the determinant of the spacial metric and the index $i$ stands for the space component. 
Therefore, the volume integral of the left-hand side of Eq. (\ref{surface}) or (\ref{surface2}) becomes 
the surface integral and then it is evaluated as 
%
\begin{eqnarray}
\int_{S_\infty}\Bigl(\partial_i V^2 -Vr_i \Bigr)dS^i =\omega_{n-2}r^{n-2}
\Bigl(\partial_r V^2+\frac{4}{(n-2)(n-1)}\Lambda r \Bigr) =2(n-3)\omega_{n-2}M,
\end{eqnarray} 
%
where $\omega_{n-2}$ is the volume of the unit $(n-2)$-dimensional sphere. 

For asymptotically flat spacetimes, we do not introduce $r^a$ and the evaluation at the boundary is same with 
the above in the limit of $\Lambda=0$. 

\section{Alternative proof}

For asymptotically anti-deSitter spacetimes, one may want to avoid to introduce the 
vector $r^a$ satisfying Eq. (\ref{defr}). This is because its introduction is artificial. 
Therefore, we try to present an alternative proof. Here we focus 
on the Einstein gravity with complex scalar fields. In the following proof, unfortunately, 
we cannot include the $p$-form fields. The argument here is basically following Ref. \cite{Wang02} 
which devoted to the vacuum case (See also Ref. \cite{Qing04}). 

In the absence of the $p$-form fields, the volume integrals of Eqs. (\ref{UE}) and (\ref{idE}) give us 
%
\begin{equation}
\omega_{a_1 \cdots a_{n-2}}=0
\end{equation} 
%
and then the spacetimes must be static. Here we assumed that the scalar fields is also stationary, 
$\mbox \pounds_k \pi=0$. Then we can employ the following metric form
%
\begin{equation}
ds^2=-V^2(x^i)dt^2+g_{ij}(x^k)dx^i dx^j.
\end{equation} 
%
The Ricci tensor becomes 
%
\begin{eqnarray}
& &R_{00}= VD^2V=\frac{n-1}{\ell^2}V^2+S_{00} \\
& &R_{ij}= {}^{(n-1)}R_{ij}-\frac{1}{V}D_i D_j V=-\frac{n-1}{\ell^2}g_{ij}+S_{ij},
\end{eqnarray} 
%
where $\ell^{-2}=-2\Lambda/(n-1)(n-2)$ and 
%
\begin{equation}
S_{ab}=T_{ab}-\frac{1}{n-2}g_{ab}T=\partial_a \pi \partial_b \pi^*+\partial_a \pi^* \partial_b \pi.
\end{equation} 
%
For the current case, $S_{00}=0$ and $S_{ij}=D_i \pi D_j \pi^*+D_i \pi^* D_j \pi$. 

The $(n-1)$-dimensional Ricci scalar becomes 
%
\begin{equation}
{}^{(n-1)}R=-\frac{(n-1)(n-2)}{\ell^2}+\frac{1}{V^2}S_{00}+S^i_i=-\frac{(n-1)(n-2)}{\ell^2}+2|D \pi|^2. 
\end{equation} 
%

Using the Einstein equations, we have the equation 
%
\begin{eqnarray}
D^2 \psi -V^{-1}D_i V D^i \psi 
& = & 2 \Bigl( D_i D_j V-\frac{1}{\ell^2}g_{ij}V \Bigr)^2+2S^{ij}D_i V D_j V 
+\frac{2D^i V}{V}D_i S_{00}-\frac{2}{V^2}S_{00}(DV)^2+\frac{2}{\ell^2}S_{00} \nonumber \\
& = & 2\Bigl( D_i D_j V-\frac{1}{\ell^2}g_{ij}V \Bigr)^2+  4 |D^i VD_i \pi|^2 \geq 0, \label{psi}
\end{eqnarray} 
%
where $\psi=(DV)^2+\ell^{-2}(1-V^2)$. Since $\psi \to 0$ as $r \to \infty $, the maximum principle implies 
$\psi \leq 0$. Thus, we have the following inequality 
%
\begin{eqnarray}
(DV)^2 \leq (V^2-1)\ell^{-2}.
\end{eqnarray} 
%
We will used this later soon. 

Let us perform the conformal transformation defined by 
%
\begin{eqnarray}
\bar g_{ij}=(1+V)^{-2}g_{ij}. 
\end{eqnarray} 
%
Then we see that the Ricci scalar of $\bar g_{ij}$ is non-negative, 
%
\begin{eqnarray}
{}^{(n-1)} \bar R (1+V)^{-2} & = &  {}^{(n-1)}R+2(n-2)\frac{D^2V}{V+1}-(n-1)(n-2)\frac{(DV)^2}{(1+V)^2} \nonumber \\
& = &  \Bigl(\frac{1}{V^2}+\frac{2(n-2)}{V(V+1)} \Bigr)S_{00}+S^i_i 
+\frac{(n-1)(n-2)}{\ell^2}\Bigl(-\frac{1-V}{1+V}-\ell^2 \frac{(DV)^2}{(1+V)^2} \Bigr) \nonumber \\
& = & -\frac{(n-1)(n-2)}{(1+V)^2}\psi+ 2 |D\pi|^2 \geq 0. \label{barr}
\end{eqnarray} 
%

The $r=\infty$ boundary is the unit sphere in the conformally transformed space and the 
trace of the extrinsic curvature is $\bar k|_{r=\infty}=n-2$. Thus, the conformally transformed space is 
a compact Riemannian manifold, $\bar M$, with the boundary of the unit sphere, $\partial \bar M=:S^{n-2}$. 
Pasting the flat space removing the ball with the unit radius with $\bar M$ along $S^{n-2}$, 
we can construct the manifold with the zero mass. Since the Ricci scalar is non-negative 
there, we can apply the positive mass theorem \cite{Miao02, Schoen81} 
and then see that the space should be flat. This means $\bar R=0$ and then Eq. (\ref{barr}) implies 
%
\begin{equation}
D_i \pi=0
\end{equation}
%
and
%
\begin{eqnarray}
\psi=0.
\end{eqnarray}
%
Then Eq. (\ref{psi}) implies 
%
\begin{eqnarray}
D_iD_jV=\frac{1}{\ell^2}g_{ij}V. 
\end{eqnarray} 
%
Therefore, ${}^{(n-1)}R_{ij}=-(n-2)\ell^{-2}g_{ij}$. 
Due to the conformal flatness, the Weyl tensor with respect to $g_{ij}$ is zero and then the Riemann tensor 
becomes ${}^{(n-1)}R_{ijkl}=-\frac{1}{\ell^2}(g_{ik}g_{jl}-g_{il}g_{jk}) $. Using the Einstein equations, we 
can compute the Riemann tensor of spacetime as follows
%
\begin{eqnarray}
& & R_{0i0j}=VD_iD_jV=\frac{V^2}{\ell^2}g_{ij}=-\frac{1}{\ell^2}g_{00}g_{ij} \\
& & R_{ijkl}={}^{(n-1)}R_{ijkl}=-\frac{1}{\ell^2}(g_{ik}g_{jl}-g_{il}g_{jk}).
\end{eqnarray} 
%
Therefore, $R_{abcd}=-\ell^{-2}(g_{ac}g_{bd}-g_{ad}g_{bc})$ holds and then the spacetime is exactly 
anti-deSitter spacetimes.



\end{document}